\documentstyle[twoside, epsf]{article}

\input ibvs2.sty

\begin{document}
\IBVShead{5xxx}{00 Month 200x}

\IBVStitletl{Early spectroscopy and photometry of the}{new outburst of V1647 Ori}

\IBVSauth{KUN, M.$^1$}

\IBVSinst{Konkoly Observatory, H-1525 Budapest, P.O.Box 67, Hungary, e-mail: kun@konkoly.hu}

\SIMBADobj{V1647 Ori}
\IBVSabs{Broad-band photometric and moderate-resolution spectroscopic}
\IBVSabs{observationsof the young eruptive star V1647 Ori, obtained between}
\IBVSabs{28 August and 1 September 2008, are presented. The observed magnitudes}
\IBVSabs{and emission line equivalent widths of the star indicate that the}
\IBVSabs{initial conditions of the new outburst are very similar to those of} 
\IBVSabs{the previous one in 2004.}

\begintext

V1647 Ori is a young eruptive variable star, illuminating a variable reflection 
nebula (McNeil's Nebula). The previous outburst of the star between 2004 January and 
2005 October has been extensively documented in the literature 
(e.g. Brice\~no et al. 2004, Ojha et al. 2006, Acosta-Pulido et al. 2007, Fedele et al. 2007). 
Optical, near-, and mid-infrared observations of the star during the quiescent period 
following the outburst (Aspin et al. 2008) suggested a spectral type of M0$\pm$0.2,
mass 0.8$\pm$0.2\,M$_{\odot}$, and age $<$\,0.5\,Myr for V1647 Ori. 
The observed properties of the outburst of V1647~Ori are different in several 
respects from both the EXor and FUor type outbursts, and suggest that this star
probably represents a new type of of eruptive young stars, younger and more deeply embedded 
than EXors, and exhibiting variations on shorter time scales than FUors.

A new outburst of the star was announced on 27th August 2008.  Itagaki (2008) 
detected the apparent brightening of V1647~Ori 26 August. The flux-calibrated 
optical spectrum of the star, obtained by Aspin (2008) on Aug 30, showed 
strong H$\alpha$ emission line with P Cygni profile, the CaII triplet lines 
in emission, and suggested a Johnson R magnitude of 17.3, corresponding to a 
brightening of some 6~mag in the R-band with respect to the quiescent phase.

In order to compare the 
present brightness and emission line strengths of the star with those observed at the 
beginning of the outburst in 2004 I observed V1647 Ori between 28 August and 1 September 2008,
using the CAFOS instrument on the 2.2-m telescope of Calar Alto Observatory (Spain). 
Spectra covering the wavelength region of 4800--7800~\AA \ were obtained 
on  Aug  29 and 31, using the grism G--100 whose dispersion is 2.12~\AA/pix. 
Grism R--100, having a dispersion of 2.04~\AA/pix  was used 
for observing the spectral region 5800--9000~\AA \ on Aug 28, 30, and Sep 1.
The exposure time was 1800~s for each spectrum. The spectrum of a He--Hg--Rb 
lamp was observed for wavelength calibration. 
Direct images, each with an exposure time of 60~s, 
were taken immediately after the spectroscopic observations, utilizing the central 
1024$\times$1024 pixel region of the SITe 2048$\times$2048 chip. 
The image scale was 0.53$^{\prime\prime}$/pix. The 9-arcmin field of view 
included seven secondary standard stars published by Semkov (2004).  
Two {\it I\/}$_{\mathrm C}$-band images were obtained on both August 28 and 29, and two 
images both in the {\it I\/}$_{\mathrm C}$ and {\it R\/}$_{\mathrm C}$ 
bands were taken on the remaining three 
nights. Data reduction and analysis were performed in IRAF. One-dimensional 
spectra were extracted from the spectroscopic images using the 
`apextract' package of IRAF. The spectrum of the nebula was also 
extracted from the images obtained through grism G--100. The resulting 
spectra were wavelength-calibrated and analysed in the `onedspec' package 
of IRAF. The direct images, obtained through the same filter on each night, 
were coadded after bias subtraction and flatfield correction, using dome 
flat field images. The instrumental  magnitudes of V1647~Ori and 
the comparison stars were determined on the coadded images by PSF-photometry 
using the `daophot' package in IRAF. The preliminary aperture photometry, 
used for scaling the PSF magnitudes was obtained using 1.5~arcsec apertures 
in each image. The instrumental 
magnitudes were transformed into the standard photometric 
system as described by Acosta-Pulido et al. (2007). 

The upper panel of Fig.~1 shows the spectrum of the star (thick line) 
and the nebula (thin line) in the green spectral region, normalized to the continuum. 
The most conspicuous feature of this spectral region, is the strong H$\alpha$ emission with 
P Cygni type absorption. Both spectra indicate a weak H$\beta$ and NaI\,D absorption. 
The lower panel shows the red spectral region with strong H$\alpha$ and CaII triplet
emissions. In addition to the strong atmospheric absorption bands around 6860, 7600, 
and 8280\,\AA \ the OI line at 7773\,\AA \ is clearly seen in absorption in each 
red spectrum. 
The left panel of Fig.~2 shows an I-band image, centred on V1647 Ori, obtained
on 1~September. The right panel shows  the H$\alpha$ line observed on three different
nights.

The  {\it R\/}$_{\mathrm C}$ and  {\it I\/}$_{\mathrm C}$  magnitudes, as well as the 
equivalent widths of the H$\alpha$, CaII, and OI lines are listed in Table~1. 
Values for both the emission and absorption components of the
H$\alpha$ line are shown. The UT, given in Column~2, refers to 
the start of the spectroscopic exposure. The photometric uncertainties were 
computed as the quadratic sum of the formal errors of the instrumental 
magnitudes provided by IRAF and the uncertainties of the standard transformation.
The uncertainties of the equivalent widths are around 6\%, estimated from  
repeated measurements. 

\vskip 0.5cm

{\samepage
\centerline{Table 1. Results of the observations}
\vskip 3mm
\begin{center}
\begin{tabular}{lcc@{\hskip2mm}cc@{\hskip2mm}c@{\hskip2mm}c@{\hskip2mm}c@{\hskip2mm}cc}
\hline
\noalign{\smallskip}
Date & UT &  {\it R\/}$_{\mathrm C}$  & {\it I\/}$_{\mathrm C}$ & \multicolumn{2}{c}{W(H$\alpha$)}   
& \multicolumn{3}{c}{W$_{\lambda}$(CaII)} & OI   \\
  &   &   &  &  em. & abs. & (8498)  & (8542) &  (8662) \\
 & & (mag) & (mag) & (\AA) & (\AA) & (\AA) & (\AA) & (\AA) & (\AA) \\
\noalign{\smallskip}
\hline
\noalign{\smallskip}
2008 Aug 28 & 04:17 & $\cdots$      & 14.64\,(0.06) & $-$31.5 & 1.6 & $-$7.46 & $-$7.84 & $-$6.24       & 1.6  \\
2008 Aug 29 & 03:39 & $\cdots$      & 14.80\,(0.07) & $-$41.5 & 3.4 & $\cdots$   & $\cdots$  & $\cdots$ &   \\
2008 Aug 30 & 03:48 & 17.02\,(0.07) & 14.64\,(0.05) & $-$41.3 & 5.0 & $-$8.45 & $-$8.50 & $-$6.79       & 2.2  \\
2008 Aug 31 & 03:32 & 16.81\,(0.07) & 14.66\,(0.07) & $-$41.5 & 6.0 & $\cdots$  &$\cdots$   & $\cdots$  &   \\
2008 Sep 01 & 03:51 & 17.11\,(0.05) & 14.69\,(0.04) & $-$43.6 & 3.6 & $-$7.97 & $-$8.52 & $-$6.36       & 2.0  \\
\noalign{\smallskip}
\hline
\end{tabular}
\end{center}}

\IBVSfig{8cm}{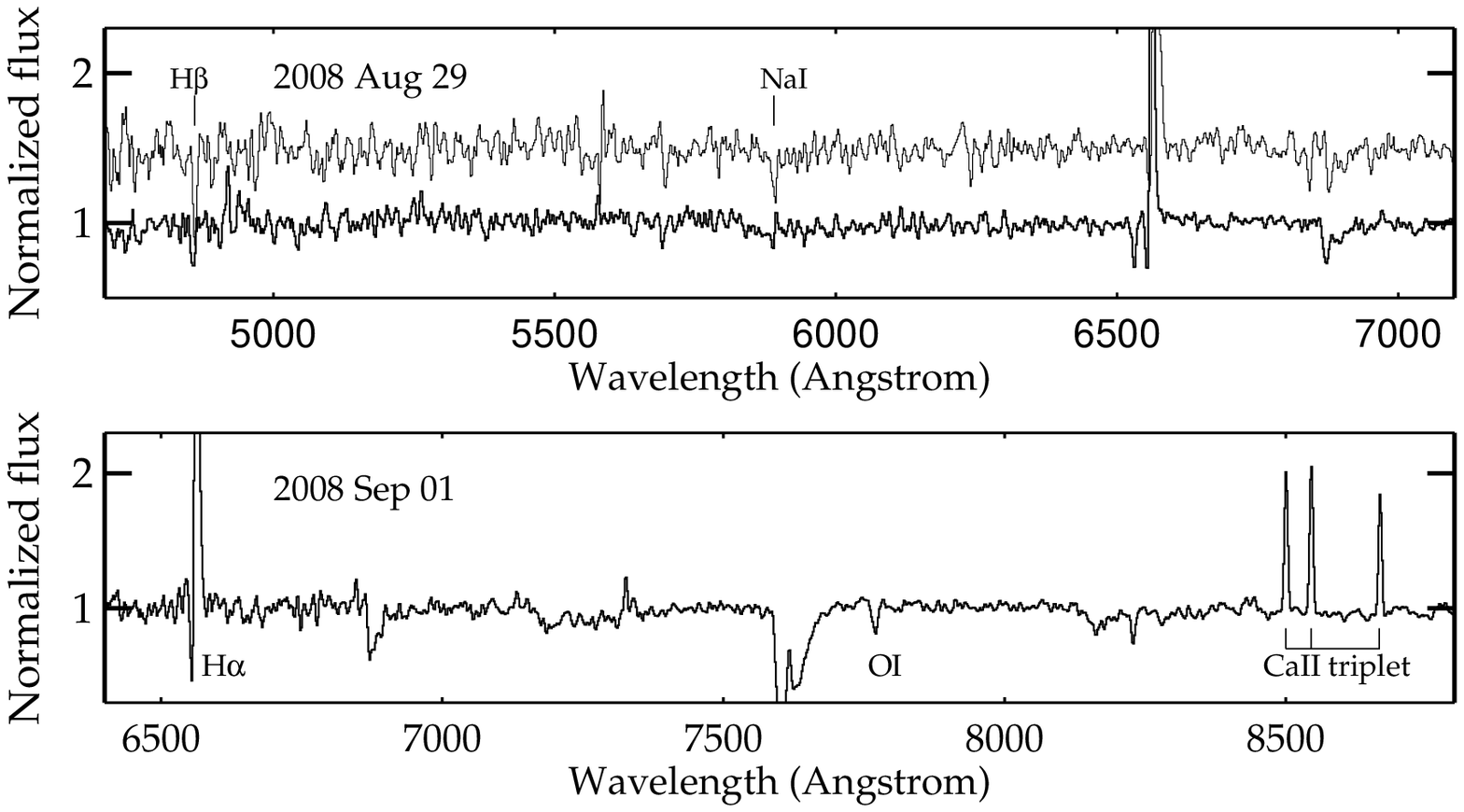}{
Upper panel: The spectrum of V1647 Ori (thick line) and McNeil's 
Nebula (thin line), on the wavelength interval 4700--7100\,\AA, obtained on Aug 29. 
Lower panel: The 6400--8800\,\AA \ region of the spectrum of V1647 Ori obtained on
Sep 01 2008.}

\IBVSfig{8cm}{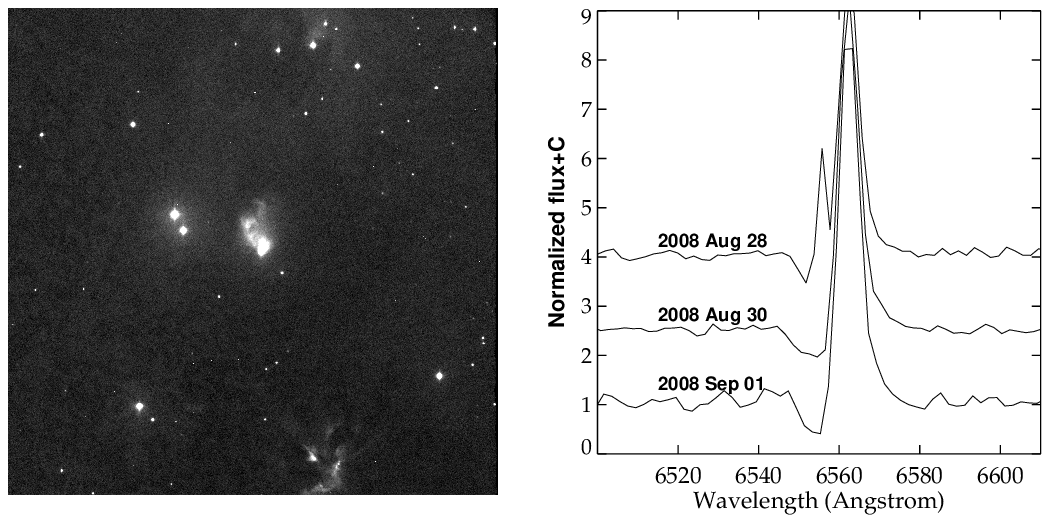}{
Left: I-band image of the field centered on V1647 Ori, observed on
September 1. Right: H$\alpha$ line profiles on three different nights.}

Comparison of the image of McNeil's Nebula, the main apparent features of the spectrum
of V1647~Ori, and the data listed in Table~1 with similar data obtained in February--March 
2004 suggests that the initial conditions of the present outburst are largely the same as 
were in 2004. Walter et al. (2004) and Ojha et al. (2006) measured similar H$\alpha$ and 
CaII equivalent widths in February--March 2004, McGehee et al.(2004), Ojha et al. (2006), 
and Acosta-Pulido et al. (2007) report similar {\it R\/}$_{\mathrm C}$ and  
{\it I\/}$_{\mathrm C}$ magnitudes for the same period.
The optical spectra obtained in February--March 2004 by Fedele et al. (2007) show
P Cygni-type profile of the H$\beta$ line with strong absorption and weak emission
component, indicating that the source of the line is a strong stellar wind. Only the absorption 
component of the H$\beta$ can be identified in the low S/N part of our spectra. 
The OI\,$\lambda$\,7773\,\AA\ absorption was also detected by Ojha et al. (2006) in the
spectra obtained at the early phases of the outburst in 2004. 
The CaII line ratios W$_{\lambda}$(8498)/W$_{\lambda}$(8542) and 
W$_{\lambda}$(8662)/W$_{\lambda}$(8542) are useful tracers of 
the physical conditions at the origin of the emission (e.g. Hamann \& Persson
1992). Our measured ratios, averaged for the three nights, are 0.95 and 0.78, whereas the
same ratios obtained by Walter et al. (2004) in  March--April 2004 are 1.10 and 0.63, 
respectively. Both measurements show that the $\lambda$\,8662\,\AA \ line 
was the weakest component of the triplet. Ojha et al. (2006) reported on nearly equal
equivalent widths of the triplet components in later phases of the outburst. The ratios 
measured in the quiescent phase by Aspin et al. (2008), 0.86 and 0.97, show a similar
situation. 
 
The simultaneous spectroscopic and photometric observations allow us to calculate line fluxes.
The average observed fluxes of the H$\alpha$, CaII($\lambda$8498), CaII($\lambda$8542), and 
CaII($\lambda$8662) emission lines are  
F(H$\alpha$)=$1.4\times10^{-17}$\,Wm$^{-2}$, F$_{\lambda}$(8498)=$1.2\times10^{-17}$\,Wm$^{-2}$,
F$_{\lambda}$(8542)=$1.3\times10^{-17}$\,Wm$^{-2}$, and 
F$_{\lambda}$(8662)=$1.0\times10^{-17}$\,Wm$^{-2}$, respectively. These numbers indicate a 
14-fold increment of the H$\alpha$ flux with respect to the flux measured in 
2007 February (Aspin et al. 2008).  The observed emission line fluxes are affected
by the increased accretion rate from the disk onto the star, the strong wind 
accompanying the enhanced accretion,  and the decreasing circumstellar 
extinction associated with the outburst (Aspin et al. 2008). The contribution of
these processes to the fluxes of various emission lines may be strongly different.

{\bf Acknowledgements:} These results are based on observations obtained at the Centro 
Astron\'omico Hispano Alem\'an (CAHA) at Calar Alto, operated jointly by the 
Max-Planck-Institut f\"ur Astronomie and the Instituto de 
Astrof\'{\i}sica de Andaluc\'{\i}a (CSIC). The observations were supported 
by the OPTICON project. OPTICON has received research funding from the European 
Community's Sixth Framework Programme under contract number RII3-CT-001566.
Financial support from the Hungarian OTKA grant T49082 is acknowledged.

\references

Acosta-Pulido, J. A., Kun, M., \'Abrah\'am, P. et al. 2007, {\it AJ}, {\bf 133}, 2020 

Aspin, C. A. 2008, {\it IAUC}, {\bf 8969}

Aspin, C. A., Beck, T. L., Reipurth, B. 2008, {\it AJ}, {\bf 135}, 423

Brice\~no, C., Vivas, A. K., Hern\'andez, J., et al. 2004, {\it ApJ}, {\bf 606}, L123

Fedele, D., van den Ancker, M. E., Petr-Gotzens, M. G., Rafanelli, P. 2007, {\it A\&A}, {\bf 472}, 207

Hamann, F., \& Persson, S. E. 1992, {\it ApJS}, {\bf 82}, 247

Itakagi, K. 2008, {\it IAUC}, {\bf 8968}

McGehee, P. M., Smith, J. A., Henden, A. A., et al. 2004, {\it ApJ}, {\bf 616}, 1058

Ojha, D.K.,  Ghosh, S. K., Tej, A. et al. 2006, {\it MNRAS}, {\bf 368}, 825

Semkov, E. H. 2004, {\it IBVS} {\bf 5578}

Walter, F. M., Stringfellow, G. S., Sherry, W. H., Field-Pollatou, A. 2004, {\it AJ}, {\bf 128}, 1872

\endreferences

\end{document}